\documentclass[tightenlines,showpacs,byrevtex,11pt,onecolumn,pra]{revtex4}%
\usepackage{graphicx}
\usepackage{dcolumn}
\usepackage{bm}
\usepackage{amsmath}
\usepackage{amsfonts}
\usepackage{amssymb}%

\begin{document}
\preprint{v65}
\title{Quantum Positioning System}
\author{Thomas B. Bahder}
\affiliation{U. S. Army Research Laboratory\\
2800 Powder Mill Road\\
Adelphi, Maryland 20783-1197}
\email{bahder@arl.army.mil}
\date{June 3, 2004}

\begin{abstract}
A quantum positioning system (QPS) is proposed that can provide a user with
all four of his space-time coordinates. The user must carry a corner cube
reflector, a good clock, and have a two-way classical channel of communication
with the origin of the reference frame. Four pairs of entangled photons
(biphotons) are sent through four interferometers: three interferometers are
used to determine the user's spatial position, and an additional
interferometer is used to synchronize the user's clock to coordinate time in
the reference frame. The spatial positioning part of the QPS is similar to a
classical time-of-arrival \ (TOA) system, \ however, a classical TOA system
(such as GPS) must have synchronized clocks that keep coordinate time and
therefore the clocks must have long-term stability, whereas in the QPS only a
photon coincidence counter is needed and the clocks need only have short-term
stability. Several scenarios are considered for a QPS: one is a terrestrial
system and another is a space-based-system composed of low-Earth orbit
(LEO)\ satellites. \ Calculations indicate that for a space-based system,
neglecting atmospheric effects, a position accuracy below the 1 cm-level is
possible for much of the region near the Earth. \ The QPS may be used as a
primary system to define a global 4-dimensional reference frame.

\end{abstract}

\maketitle

\section{Introduction}

During the past several years, the Global Positioning System (GPS) has
practically become a household word. \ The GPS is a U.S. Department of Defense
satellite system that is used by both the military and civilians for
navigation and time
dissemination\cite{ParkinsonSpilker1996,Kaplan96,Hofmann-Wellenhof93}.
\ Automobile, ship, aircraft, and spacecraft use the GPS for navigation.
Telephone and computer network systems that require precise time use the GPS
for time synchronization. \ The GPS is a complex system consisting\ of
approximately 24 satellites orbiting the Earth in circular orbits at
approximately 4.25 Earth radii. \ The GPS is designed so that signals travel
line-of-site from satellite to user, and from any place on Earth at least four
satellites are in view. \ If a user receives four GPS satellite signals
simultaneously from four satellites,\ $s=1,2,3,4$, \ and the satellites'
space-time coordinates $(t_{s},\mathbf{r}_{s})$ at time of transmission are
known,\ the user can solve for his unknown space-time coordinates,
$(t_{o},\mathbf{r}_{o})$, by solving the four
equations\cite{Bahder2001,Bahder2003}%

\begin{equation}
|\mathbf{r}_{o}-\mathbf{r}_{s}|^{2}-c^{2}(t_{o}-t_{s})^{2}%
=0,\;\;\;\;s=1,...4\label{GPSlightcones}%
\end{equation}
In Eq.(\ref{GPSlightcones}) we assume that the signals travel on four light
cones that are centered at the reception event and we have ignored atmospheric
delays. \ \ The actual signals that the GPS satellites transmit are
continuous-wave circularly polarized radio-frequency signals on two carrier
frequencies in the L-band centered about: $L_{1}\approx1575.42$ MHz and
$L_{2}\approx1227.6$ MHz. The carrier frequency signals are modulated by a
pseudorandom noise (PRN) code. \ A GPS\ receiver makes a phase difference
measurement, called a \textit{pseudorange measurement}, which is the phase
difference between the PRN code received from the satellite and an identical
copy of the PRN code that is replicated inside the GPS\ receiver,\ see
Ref.\cite{Bahder2003} for details of the GPS pseudorange measurement process.
\ The pseudorange measurement is essentially made by performing a correlation
of the code bits in the PRN code received from the satellite with an identical
copy of the same PRN code replicated inside the GPS receiver. \

Recently, there have been several proposals for synchronizing clocks by making
use of entangled quantum
systems\cite{Chuang2000,Jozsa2000,Giovannetti2001,Giovannetti2001b,Giovannetti2002,Bahder2004,Shih2003}%
. \ The question naturally arises whether entangled quantum systems can be
exploited to determine all four space-time coordinates of a user, rather than
just time.

In this article, I describe a quantum positioning system (QPS) that is
in-principle capable of providing a user with all four of his space-time
coordinates. This QPS\ is the quantum analog of the classical GPS described
above. \ The QPS is based on entangled photon pairs (biphotons) and second
order correlations within each pair. \ The two-photon coincidence counting
\ rate is the basic measured quantity. \ In order to determine his four
space-time coordinates, a user of the QPS must carry a corner cube reflector,
a good clock, and have a two-way classical channel for communication with the
origin of the reference frame.

\section{Interferometric Quantum Positioning System}

For simplicity of discussion, I assume that space-time is flat with
Minkowski\cite{Bahder2003} coordinates $(t,x,y,z)$ and that the user of the
QPS is stationary. \ The complete QPS consists of four biphoton sources
(entangled photon pairs), four beam splitters and four two-photon coincidence
counting Hong-Ou-Mandel (HOM) interferometers, see Figure 1. \ Three of the
interferometers are used together to solve for the user's position and one
interferometer is used to solve for the user's time, in a particular reference frame.

Six spatial points, \ $\mathbf{R}_{i}$, where $\mathbf{R}_{i}=(x_{i}%
,y_{i},z_{i})$, for $i=1,2,3,...,6$, define the spatial part of the reference
frame at constant coordinate time $t$. \ The six points\ $\mathbf{R}_{i}$
define three independent baselines in pairs, $(\mathbf{R}_{1},\mathbf{R}_{2}%
)$, $(\mathbf{R}_{3},\mathbf{R}_{4})$, and $(\mathbf{R}_{5},\mathbf{R}_{6})$.
\ \ The points \ $\mathbf{R}_{i}$ are assumed to be accurately surveyed, so
their coordinates are precisely known. Determination of a user's spatial
coordinates,\ $\mathbf{r}_{o}=(x_{o},y_{o},z_{o})$, is done with respect to
this reference frame.\ \ A stationary clock in this reference frame, say at
the origin of coordinates, $(x,y,z)=(0,0,0)$, provides a measure of coordinate
time $t$ in this 4-dimensional system of coordinates. \ We neglect all
gravitational effects\cite{Bahder2001} so that the user's clock, which keeps
proper time $\tau$, runs at the same rate as coordinate time $t$ in the
reference frame defined by the spatial points $\mathbf{R}_{i}$, so that
$d\tau/dt=1$\ . \ Synchronization of the user's clock to coordinate
time\cite{Bahder2004b} means that we have a method to compute the integration
constant $\tau_{o}$ in $\tau-\tau_{o}=t$. \ \ In four dimensional flat
space-time, the world lines of the spatial points $\mathbf{R}_{i}$ define a
tube. \ Events that are simultaneous in this system of 4-dimensional
coordinates occur at constant values of coordinate time $t$, which is a
hyperplane that cuts this tube. \ \

There are several possible modes for the QPS. \ First, consider a user that
must determine all four of his space-time coordinates $(t_{o},x_{o}%
,y_{o},z_{o})$. \ In a previous paper\cite{Bahder2004}, an algorithm has been
given to synchronize the proper time $\tau$ on a user's clock to coordinate
time, $t$, without prior knowledge of the geometric range from the reference
clock (which keeps coordinate time $t$) to the user clock. \ We assume that
this algorithm is employed here to synchronize the user's clock with
coordinate time $t$. \ This algorithm requires a two-way classical channel of
communication between the user and the reference frame origin, where
coordinate time $t$ is kept.\ \ The three spatial coordinates are determined
separately as follows, refer to Figure 1. \

Each baseline, such as the one connected by points, $(\mathbf{R}%
_{1},\mathbf{R}_{2})$, contains an entangled photon pair (biphoton)
source\cite{Burnham1970,HOM1987,Ghosh1986,KlyshkoBook1988,Rubin1994} located
in the baseline. \ For convenience, we take the biphoton source to be at the
midpoint of the baseline at point $E_{1}$ at position $\mathbf{r}_{1}%
$.\ \ Additionally, each baseline contains a 50:50 beam splitter and two
photon detectors, see Figure \ref{boxE1}. \ For simplicity of discussion, we
assume that the biphoton source is essentially collocated with the beam
splitter and two photon detectors at point $E_{1}$, see Figure 2. \ Along
\ the baseline there is a controllable optical delay at point $D_{1}$. The
other two baselines contain the same equipment, as shown in Figure 2. \ The
QPS works as follows. \ \ \

Photon pairs are created at $E_{1}$, are sent to positions $\mathbf{R}_{1}%
$\textbf{\ }and\textbf{\ }$\mathbf{R}_{2}$, and are re-directed to the user at
the unknown position \ $\mathbf{r}_{o}$. \ \ The two photon paths are similar,
except that one path has a controllable optical delay $D_{1}$. The optical
delay is assumed to be calibrated so that we can accurately impose an
arbitrary delay time\cite{delaytime}. Next, the entangled photons reflect from
the user's corner cube reflector at $\mathbf{r}_{o}$, and return back through
the same paths, through points $\mathbf{R}_{1}$\textbf{\ }and\textbf{\ }%
$\mathbf{R}_{2}$, and arrive at a HOM interferometer that is collocated at
$E_{1}$ at position $\mathbf{r}_{1}$, see Figure \ref{boxE1}. For convenience,
we assume that the interferometer is collocated with the photon generation
point $\mathbf{r}_{1}$. Again, both photon return paths are similar, but one
path has the optical delay $D_{1}$. We have the following effective round-trip
times for each photon path%
\begin{align}
t_{L}  &  =\frac{2}{c}[|\mathbf{r}_{o}-\mathbf{R}_{1}|+|\mathbf{R}%
_{1}-\mathbf{r}_{1}|]\label{propagation times}\\
t_{R}  &  =\frac{2}{c}[|\mathbf{r}_{o}-\mathbf{R}_{2}|+|\mathbf{r}%
_{1}-\mathbf{R}_{2}|+(n-1)\ d]\nonumber
\end{align}
where $d$ is the geometric thickness of the optical delay $D_{1}$
perpendicular to the optical path and $n$ is the effective index of refraction
for the optical delay $D_{1}$. \ The optical delay $D_{1}$ is now adjusted
until a minimum is observed in the two-photon counting rate at $E_{1}%
$\cite{HOM1987}. \ When the minimum in photon coincidence counting rate is
observed at interferometer $E_{1}$, the effective travel times for each photon
path are equal, $t_{L}=t_{R}$. The interferometer is balanced when the
condition $t_{L}=t_{R}$ is satisfied, and a unique minimum is observed in the
two-photon counting rate $R_{c}$. The accuracy with which this minimum can be
observed depends on the bandwidth $\Delta\omega$ of the band-pass interference
filters used in front of the photon detectors.

We get an equation that relates the geometric path lengths to the measured
optical delay time $\Delta t_{1}=(n-1)d/c$:%
\begin{equation}
|\mathbf{r}_{o}-\mathbf{R}_{1}|+|\mathbf{R}_{1}-\mathbf{r}_{1}|=|\mathbf{r}%
_{o}-\mathbf{R}_{2}|+|\mathbf{r}_{1}-\mathbf{R}_{2}|+c\Delta t_{1}%
\label{leg11}%
\end{equation}
\ \ An analogous measurement process is carried out for the other two
baselines. \ For simplicity,\ I assume that the points $E_{1}$, $E_{2}$, and
$E_{3}$ are located at midpoints of their baselines. \ We then obtain the
three equations%
\begin{equation}
|\mathbf{r}_{o}-\mathbf{R}_{1}|=|\mathbf{r}_{o}-\mathbf{R}_{2}|+s_{1}%
\label{baseline 1}%
\end{equation}%
\begin{equation}
|\mathbf{r}_{o}-\mathbf{R}_{3}|=|\mathbf{r}_{o}-\mathbf{R}_{4}|+s_{2}%
\label{baseline 2}%
\end{equation}%
\begin{equation}
|\mathbf{r}_{o}-\mathbf{R}_{5}|=|\mathbf{r}_{o}-\mathbf{R}_{6}|+s_{3}%
\label{baseline 3}%
\end{equation}
where $s_{i}=c\Delta t_{i}$ for $i=1,2,3$. \

The two-photon coincidence counting rate is given
by\cite{Glauber1965,Glauber1963,Mandel1995}%
\begin{equation}
R_{c}=\alpha_{1}\alpha_{2}|\eta V|^{2}|G(0)|^{2}[1-e^{-(\Delta\omega\Delta
t_{1})^{2}}]\label{coincidence count rate}%
\end{equation}
where $|V|^{2}$ is the pump intensity in photons per second, $\alpha_{1}$ and
$\alpha_{2}$ are the quantum efficiencies of detectors $D_{1}$ and $D_{2} $,
$\eta$ is a dimensionless constant and $G(t)$ is the Fourier transform of the
spectral function $\phi$, which is the auotocorrelation function of the
down-converted light
\begin{equation}
G(t)=%
{\displaystyle\int\limits_{0}^{\infty}}
\phi(\frac{1}{2}\omega_{0}+\omega,\frac{1}{2}\omega_{0}-\omega)e^{-i\omega
t}d\omega\label{G2 function}%
\end{equation}

The Eqs.(\ref{baseline 1})-(\ref{baseline 3}) can be solved for the three
unknown user spatial coordinates, $\mathbf{r}_{o}=(x_{o},y_{o},z_{o})$, in
terms of the three measured time delays, $\Delta t_{1}$, $\Delta t_{2}$,
$\Delta t_{3}$, which balanced the three interferometers. \ The measured data
consists of photon coincidence count rate vs. optical time delay lengths
$s_{i}$, for $i=1,2,3$. \ \ Clearly a search must be done of the data to
locate the minimum that corresponds to equal time of travel along the
interferometer arms. \ The computations can be done at points $E_{1}$, $E_{2}
$, and $E_{3}$. \ This search to locate the minimum is the quantum analog of
the correlation of the PRN\ code in a classical GPS\ receiver, which was
described above.\ \ \ When the three interferometers at $E_{1}$, $E_{2}$, and
$E_{3}$ have been balanced simultaneously, a classical message is sent to the
user giving him the values of his coordinates $\mathbf{r}_{o}=(x_{o}%
,y_{o},z_{o})$. Clearly, classical communication is needed between the points
$\mathbf{R}_{1}$, $\mathbf{R}_{2}$, and $\mathbf{R}_{3}$ to establish that the
interferometers are balanced at a given coordinate time $t$. We imagine that
when each interferometer is balanced, a message is sent to the origin of
coordinates. \ When three messages are simultaneously received at the origin
of coordinates (saying that the three interferometers are balanced),
Eqs.(\ref{baseline 1})-(\ref{baseline 3}) are solved for the user's
coordinates $\mathbf{r}_{o}=(x_{o},y_{o},z_{o})$ and the user's coordinates
are sent to the user through a classical channel of communication.

In the QPS that we describe, there is an apparent asymmetry in the
determination of a user's spatial coordinates, $\mathbf{r}_{o}=(x_{o}%
,y_{o},z_{o})$, and in the determination of the user's time. \ In my view,
this asymmetry is a reflection of the asymmetric way that space and time enter
in the theory of the quantized electromagnetic field to give rise to photons
as quanta of the field. \ As mentioned above, the time synchronization of the
user's clock to coordinate time is done by a method previously described by
Bahder and Golding\cite{Bahder2004}. \ Therefore in what follows, I discuss
only the spatial part of the QPS.

With some modification of the above scheme, we may imagine that we could
design a similar system based on first-order coherence for position
determination\cite{Glauber1965,Glauber1963}. A single beam from a
\ continuous-wave laser can be split and the beams sent on two different
paths. However, in such a case, there would be an ambiguity that is associated
with the wavelength of the light (interference fringes will be seen) that is
unresolvable in principle. \ In contrast, in the quantum case (which relies on
second-order coherence) the ambiguity is resolved because equal propagation
times for two paths lead to quantum interference: equal travel times for two
paths create a \textit{unique} observable minimum \ in the two-photon
coincidence counting rate $R_{c}$.\

The measured quantities in the QPS are the optical path delays $s_{i}$. \ For
a given measured value of optical delay, say $s_{1}$, Eq.(\ref{baseline 1})
specifies that the user's coordinates must lie on a hyperboloid surface with
foci at $\mathbf{R}_{1}$ and $\mathbf{R}_{2}$, i.e., a hyperbola of revolution
that is symmetric about the baseline defined by $\mathbf{R}_{1}$ and
$\mathbf{R}_{2}$. \ The user's position, $\mathbf{r}_{o}$, is then given by
the intersection of three hyperbolas given by Eq.(\ref{baseline 1}%
)-(\ref{baseline 3}). \ Each Eq.(\ref{baseline 1})-(\ref{baseline 3}) is just
the equation for a baseline in a classical time of arrival (TOA)\ system that
records arrival times of classical light pulses (or distinct intensity edges)
at two spatial reception points $\mathbf{R}_{i}$. \ In the case of a classical
TOA\ system, pulse\ arrival time at \textit{four} locations \ is needed to
determine all four space-time coordinates. In that case, four time difference
of arrival (TDOA) equations can be formed from four points, and each point is
used multiply to (effectively) form the baselines.\ (Taking TDOAs results in a
system of three equations where the emission event time has cancelled out. )
\ In the quantum case, since correlations between photon pairs are used, we
must use three baselines defined by six points $\mathbf{R}_{i}$, plus an
additional interferometer for the determination of the user's time. \ As we
will see below, the QPS\ is an interferometric system. \

More fundamentally, \ and more significant for applications, is that in the
classical case of a TOA system, we must have good clocks that are synchronized
to coordinate time so that accurate pulse arrival times at the four
$\mathbf{R}_{i}$ reception points can be recorded. A good clock that keeps
coordinate time is often a difficult requirement to meet in
practice\cite{clock stability}. \ In contrast, in the quantum case two-photon
coincidence counts at detectors $D_{1}$and $D_{2}$ are measured and only a
good "flywheel" clock is needed (i.e., a clock having a good short-term
stability) to measure photon coincidence count rates while the optical time
delay is adjusted, to locate the minimum in $R_{c}$.

\section{\textbf{Geometric Dilution of Precision}}

In the case of the classical GPS, the geometrical positions of the GPS
satellites determine the accuracy of the user's position. \ \ This effect is
sometimes called the geometric dilution of precision (GDOP). We compute the
positioning accuracy and the effect of GDOP for the QPS from
Eq.(\ref{baseline 1})-(\ref{baseline 3}). $\ \ $These\ equations give an
implicit relation $\mathbf{r}_{o}=\mathbf{r}_{o}(\mathbf{R}_{1},\mathbf{R}%
_{2},\mathbf{R}_{3},\mathbf{R}_{4},\mathbf{R}_{5},\mathbf{R}_{6},s_{1}%
,s_{2},s_{3})$ for the user position $\mathbf{r}_{o}$ as a function of the
three measured path delays, $s_{i}$, and the six baseline endpoints
$\mathbf{R}_{i}$. If we knew the error in the measured path length delays,
$ds_{1}$, $ds_{2},$and $ds_{3}$, we could compute the error in the three
components of the user's position vector, $d\mathbf{r}_{o}=(dx_{o}%
,dy_{o},dz_{o},)$, from
\begin{equation}
d\mathbf{r}_{o}=%
{\displaystyle\sum\limits_{i=1}^{3}}
\frac{\partial\mathbf{r}_{o}}{\partial s_{i}}ds_{i}\label{position error}%
\end{equation}
for constant $\mathbf{R}_{i}$. \ However, these errors are statistical in
nature, so instead I compute the standard deviations $\sigma_{x}$, $\sigma
_{y}$, and $\sigma_{z}$, of the user coordinates $x_{o}$, $y_{o}$, and $z_{o}
$,\ as a function of the standard deviations $\sigma_{s_{1}}$, $\sigma_{s_{2}%
}$, and $\sigma_{s_{3}}$, of the measured optical time delays $s_{1}$, $s_{2}%
$, and $s_{3}$. \ For constant $\mathbf{R}_{k}$ for $k=1,...,6$, these
standard deviations are related by\cite{Bevington2003}%
\begin{align}
\sigma_{x}^{2}  &  =\left(  \frac{\partial x_{o}}{\partial s_{1}}\right)
^{2}\sigma_{s_{1}}^{2}+\left(  \frac{\partial x_{o}}{\partial s_{2}}\right)
^{2}\sigma_{s_{2}}^{2}+\left(  \frac{\partial x_{o}}{\partial s_{3}}\right)
^{2}\sigma_{s_{3}}^{2}\label{xyz standard deviations}\\
\sigma_{y}^{2}  &  =\left(  \frac{\partial y_{o}}{\partial s_{1}}\right)
^{2}\sigma_{s_{1}}^{2}+\left(  \frac{\partial y_{o}}{\partial s_{2}}\right)
^{2}\sigma_{s_{2}}^{2}+\left(  \frac{\partial y_{o}}{\partial s_{3}}\right)
^{2}\sigma_{s_{3}}^{2}\nonumber\\
\sigma_{z}^{2}  &  =\left(  \frac{\partial z_{o}}{\partial s_{1}}\right)
^{2}\sigma_{s_{1}}^{2}+\left(  \frac{\partial z_{o}}{\partial s_{2}}\right)
^{2}\sigma_{s_{2}}^{2}+\left(  \frac{\partial z_{o}}{\partial s_{3}}\right)
^{2}\sigma_{s_{3}}^{2}\nonumber
\end{align}
where the partial derivatives are done at\ constant $\mathbf{R}_{k}$. \ The
lengthy calculation to compute the partial derivatives in
Eq.(\ref{xyz standard deviations}) is done analytically using
\textit{Mathematica}. \ For simplicity, I assume that the error distributions
of the $s_{i}$\ are Gaussian and that the three standard deviations are equal,
$\sigma_{s1}=\sigma_{s2}=\sigma_{s3}\equiv\sigma_{s}$. For a spherically
symmetric probability distribution of 3-dimensional positions $\mathbf{r}%
_{o}=(x_{o},$ $y_{o},z_{o})$, the spherical error probable (SEP), which is the
radius $R$ within which 50\%\ of the points lie, is
related\cite{Bahder1995Rep} to the standard deviations $\sigma_{x}=\sigma
_{y}=\sigma_{z}\equiv\sigma$ by $R\cong1.538\sigma$. \ In our case, the
probability distribution of $\mathbf{r}_{o}$ is not necessarily spherical. To
approximate the SEP error metric, we compute a weighted approximation to the
SEP metric by defining $R_{xyz}\cong1.538\frac{1}{\sqrt{3}}(\sigma_{x}%
^{2}+\sigma_{y}^{2}+\sigma_{z}^{2})^{1/2}$. When the error distribution for
$\mathbf{r}_{o}$ is spherically symmetrical, the error metrics are equal:
$R_{xyz}=R$. \ I consider the effect of GDOP for two cases, one in which the
interferometer baselines are near each other, and the other case where the
baselines are well separated, which is the case with classical GPS or a
classical TOA\ system.

\subsection{Geodetic Positioning System}

First, consider \ a case where the three baselines coincide with the three
Cartesian coordinate axes of a reference frame, see Figure
\ref{Terrestrial Interferometer}. Such a case might occur when the baselines
are on the Earth, and we want to determine the position of an object with
respect to a topocentric coordinate system. For example, consider the center
of the QPS at the origin of Cartesian coordinates and an object with a corner
reflector at a range of 100$%
\operatorname{m}%
$ from the QPS,\ with coordinates $\left(  x_{o},y_{o},z_{o}\right)  =(100%
\operatorname{m}%
)(1,1,1)/\sqrt{3}$. Figure \ref{CaseAContourPlot} shows a plot of contours of
constant values of $1/R_{xyz}$ in the $x_{o}-y_{o}$ plane at $z_{o}%
=100/\sqrt{3}%
\operatorname{m}%
$, for the interferometer arm (half) length $a=2%
\operatorname{m}%
$ and error (standard deviation) in optical path $\sigma_{s}=1.0\times10^{-6}%
\operatorname{m}%
$. \ In the contour plot, the position error is $R_{xyz}=8.3%
\operatorname{cm}%
$ for $\left(  x_{o},y_{o},z_{o}\right)  =(100%
\operatorname{m}%
)(1,1,1)/\sqrt{3}$, whereas for $\left(  x_{o},y_{o},z_{o}\right)  =(30%
\operatorname{m}%
,30%
\operatorname{m}%
,100/\sqrt{3})$ the error $R_{xyz}=3.9%
\operatorname{cm}%
$, which corresponds to the upper right high-accuracy (light-shaded) region in
Figure \ref{CaseAContourPlot}. \ On the $z$-axis at $x_{o}=y_{o}=0$ and
$z_{o}=100%
\operatorname{m}%
$ the error $R_{xyz}$ is essentially infinite. \ Figure \ref{CaseAPlot1} shows
a plot of the error metric $R_{xyz}$ vs. $x_{o}$, for $y_{o}=30%
\operatorname{m}%
$ and $z_{o}=100%
\operatorname{m}%
$/$\sqrt{3}$, which corresponds to a line in Figure 4 with relatively small
error $R_{xyz}$. \ In the high-accuracy light-shaded region of Figure
\ref{CaseAContourPlot},\ \ for $x_{o}=y_{o}=30%
\operatorname{m}%
$ and $z_{o}=100%
\operatorname{m}%
$/$\sqrt{3}$, \ the dependence of the error $R_{xyz}$ on the baseline length
$2a $ is plotted in Figure \ref{CaseAPlot2}, also using $\sigma_{s}%
=1.0\times10^{-6}%
\operatorname{m}%
$. \ For a four-meter baseline, \ $2a=4%
\operatorname{m}
$, the error is just under $5%
\operatorname{cm}%
$.

Note that the position error $R_{xyz}$ depends linearly on $\sigma_{s}$, which
is the standard deviation (error) in measurement of the optical path delay
needed to obtain the minimum in two-photon coincidence counts $R_{c}$. \ The
width of this minimum depends on the interference filters in front of the
photon coincidence counting detectors as well as the pump laser
bandwidth\cite{HOM1987,Grice1997}. Depending on the experimental design, this
minimum may be measured to better\ than $\sigma_{s}=1.0\times10^{-6}%
\operatorname{m}%
$, which was used in these plots, and hence accuracies may be better than plotted.

Finally, I note that the error function $R_{xyz}$ has a very complex
dependence on user coordinates $\left(  x_{o},y_{o},z_{o}\right)  $, and as
stated earlier, the error function $R_{xyz}$ also depends critically on the
way the baselines are distributed, i.e., it depends on the six points
$\mathbf{R}_{k}$ for $k=1,...,6$, which define the baseline endpoints. In the
next example, I consider a situation where the baselines do not intersect, and
thereby the error $R_{xyz}$ is considerably smaller than for the case
considered above, even though the distances are larger.

\subsection{Satellite-based QPS}

Now assume that each point of a baseline, $\mathbf{R}_{i}$, is associated with
a different satellite, and that the spatial interferometer legs are formed
from pairs of points $(\mathbf{R}_{1},\mathbf{R}_{2})$, $(\mathbf{R}%
_{3},\mathbf{R}_{4})$, and $(\mathbf{R}_{5},\mathbf{R}_{6})$, see Figure
\ref{QPS plot}. \ Specifically, I assume that the points $\mathbf{R}_{k}%
$,\ are on low-Earth orbit (LEO) satellites. \ It may seem optimistic that a
QPS is feasible with such large baselines because single photons must be
propagated over these baselines and then reliably detected. However, recently
single photons have been propagated through the atmosphere and detected over
10 $%
\operatorname{km}%
$ $\ $distance in daylight\cite{Hughes2002}, and another study concludes that
there are no obstacles to create a single-photon quantum key distribution
system between ground and low-Earth orbiting satellites\cite{Rarity2002} .
\ Therefore, a LEO-satellite QPS\ may be possible.

As an example of the positioning errors in a QPS made from LEO satellites, I
take the baseline endpoints to be: $\mathbf{R}_{1}=(a,-b/2,0)$, $\mathbf{R}%
_{2}=(a,b/2,0)$, $\mathbf{R}_{3}=(b/2,a,0)$, $\mathbf{R}_{4}=(-b/2,a,0)$,
$\mathbf{R}_{5}=(-b/(2\sqrt{2}),-b/(2\sqrt{2}),a)$, and $\mathbf{R}%
_{6}=(b/(2\sqrt{2}),b/(2\sqrt{2}),a)$. \ A plot of this configuration is shown
in Figure \ref{QPS plot}. \ \ A contour plot of the reciprocal error function,
$1/R_{xyz}$, is shown in the $x_{o}-y_{o}$ plane for $z_{o}=R_{e}/\sqrt{3}$,
where $R_{e}=6378%
\operatorname{km}%
$ is the Earth's radius, see Figure \ref{CaseCContourPlot}. \ As an example,
in the calculations below I take the semi-major axis of the LEO satellites to
be $a=7360%
\operatorname{km}%
$ and the baseline between pairs of satellites as $b=20%
\operatorname{km}%
$. $\ $The standard deviation (error) in the measured optical delay is taken
to be $\sigma_{s}=1.0%
\operatorname{\mu m}%
$. \ \ For a user on the surface of the Earth with coordinates $\left(
x_{o},y_{o},z_{o}\right)  =(1,1,1)R_{e}/\sqrt{3}$ the error is $R_{xyz}=0.10%
\operatorname{cm}%
.$ \ For these same parameters, Figure \ref{CaseCPlot1} shows a plot of the
position error $R_{xyz}$ $\ $vs. $x_{o}$ for $y_{o}=z_{o}=R_{e}/\sqrt{3}$.
\ \ Note that over a large range of $x_{o}$-values the error remains below $1%
\operatorname{cm}%
$. \ Finally, Figure \ref{CaseCPlot2} shows a plot of the position error in
the radial direction: $R_{xyz}\left(  r_{o}/\sqrt{3},r_{o}/\sqrt{3}%
,r_{o}/\sqrt{3}\right)  $ $\ $vs. $r_{o}$ \ for the same parameters. \ On a
radial line in the (1,1,1) direction, \ the error remains below $1%
\operatorname{cm}%
$ \ up to $r_{o}=11680%
\operatorname{km}%
.$ \ However, near $1300%
\operatorname{km}%
$ the error rises steeply. \ \ This is an example of the complex dependence of
$R_{xyz}$\ on user position, which was mentioned earlier.

Clearly, the geometric positioning and layout of the baselines significantly
affects the accuracy of a user's position. Note that the terrestrial QPS
(discussed in the previous section) had a ratio of baseline length to user
position $a/r_{o}=0.02$, whereas this LEO satellite QPS has $b/a=0.003$.
\ \ By comparing the baseline layout for the terrestrial QPS and this LEO
satellite QPS, it is clear that the positioning accuracy is sensitive to the
separation and layout of the baselines, but not so sensitive to the baseline
lengths. \ Other calculations (not shown) support this conclusion. \ \

The above calculations for a satellite-based QPS\ are only meant as an example
to illustrate the magnitude of errors in position that may be achievable. A
significant amount of engineering calculations must be performed to design a
realist satellite-based QPS. \ Furthermore, real satellites \ are moving and
engineering similar to that used in the classical GPS\ would have to be done,
e.g., using Kalman filtering techniques. \ Obviously, bright sources of
entangled photons (biphotons) are needed. The calculations above suggest that
if properly engineered, a satellite-based QPS may achieve position accuracy of
objects near the Earth's surface below 1$%
\operatorname{cm}%
$. \ In these calculations, I have ignored the time delays introduced by the
atmosphere. \ However, corrections can be made for atmospheric effects using
multiple colors of photons similarly to what is done with the GPS. Perhaps one
advantage of the quantum system as compared to the classical GPS is that
entangled photons exhibit group velocity dispersion cancellation, which may be
an important factor for future engineering and design of \ a QPS
\cite{Steinberg1992,Steinberg1992a,Perina1999,Valencia2002}.

\section{Alternative Scenarios}

\subsection{Position-Only Determination}

A QPS can be designed to work in several modes, depending on the needs of the
user and the required scenario. \ In the above discussion, we have described
the case where a user of the QPS wants to determine both his spatial and time
coordinates, $(t_{o},x_{o},y_{o},z_{o})$. A second alternative is that a user
may only need to obtain their spatial coordinates, and he may not need the
correct time. In this latter case, the time synchronization portion of the
system is not needed, and the user may find their position coordinates
$(x_{o},y_{o},z_{o})$ by having only a corner reflector and a one-way
(reception only) classical channel of communication with the reference frame
origin, where the simultaneity of the three two-photon coincidence counting
rate minima is established. \

Another mode of operation of a QPS is where we want to determine the position
of an object with a corner cube reflector, such as a geostationary satellite.
\ In such a case, information on the position of the satellite, $\mathbf{r}%
_{o}=(x_{o},y_{o},z_{o})$, is only needed on the ground. \ The satellite's
position coordinates can be determined on the ground using a QPS, and only a
corner cube reflector is needed on the satellite, but no communication channel
to the satellite and no on-board clock is needed.

\subsection{User Carries QPS Receiver}

\begin{figure}
[ptb]
\begin{center}
\includegraphics[
trim=-0.385748in 1.035588in -0.554747in -1.035588in,
height=5.2615in,
width=3.749in
]%
{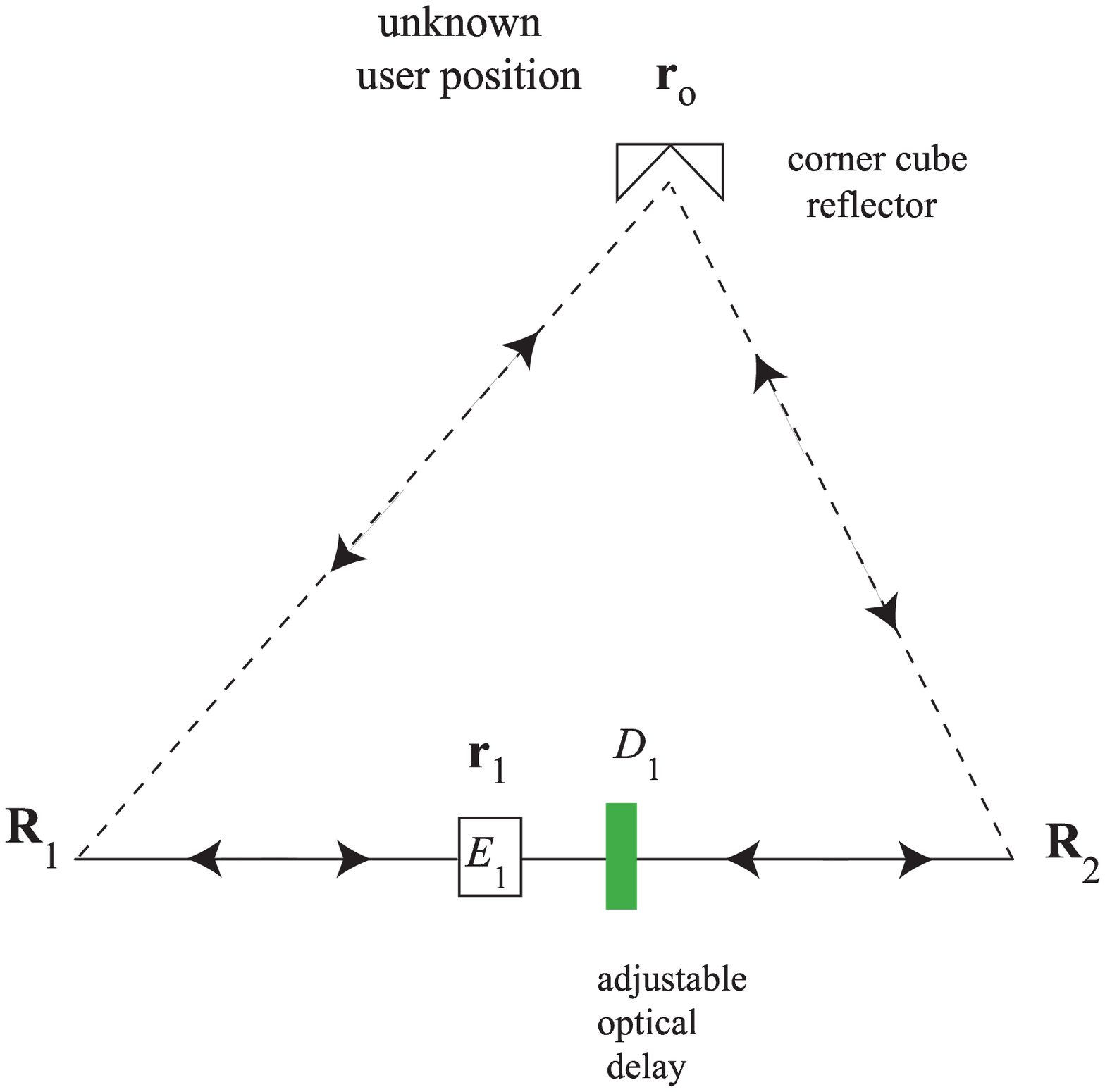}%
\caption{One baseline is shown for the quantum positioning system (QPS).
\ Points $\mathbf{R}_{1\text{ }}$ and $\mathbf{R}_{2\text{ }}$on the baseline
contribute to the definition of the reference frame for spatial positioning.
Box $E_{1}$ contains an entangled photon (biphoton) source and 50:50 beam
spliter, see Figure 2. The quantity $D_{1}$ is a controllable, calibrated
optical delay.}%
\label{baselineFigure}%
\end{center}
\end{figure}

\begin{figure}
[ptb]
\begin{center}
\includegraphics[
height=4.5472in,
width=4.964in
]%
{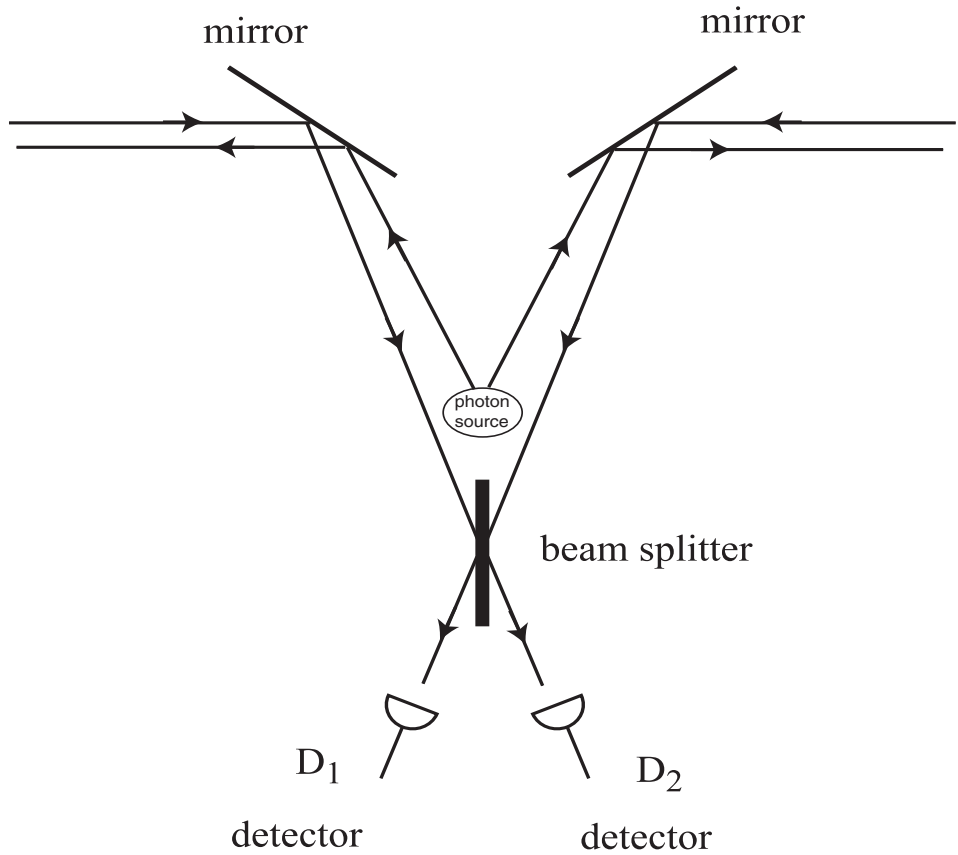}%
\caption{An expanded view of the contents of each of the three boxes $E_{1}$,
$E_{2}$, and $E_{3}$, which are located on the three baselines, one of which
is shown in Figure 1. Each box contains an entangled photon (biphoton) source,
a 50:50 beam splitter, and two single-photon detectors $D_{1}$and $D_{2}$, to
perform photon coincidence counting. }%
\label{boxE1}%
\end{center}
\end{figure}

\begin{figure}
[ptb]
\begin{center}
\includegraphics[
height=3.4264in,
width=3.8501in
]%
{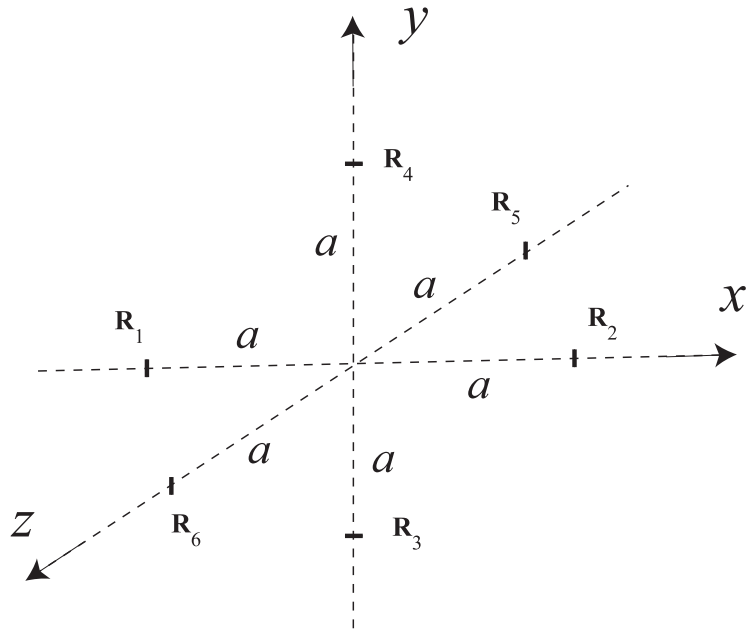}%
\caption{The baselines are shown for a possible terrestrial QPS that might be
used on the Earth. The baselines lie along the $x$, $y$, and $z$ axes, are of
length $2a$ and are orthogonal to each other.}%
\label{Terrestrial Interferometer}%
\end{center}
\end{figure}

\begin{figure}
[ptbptb]
\begin{center}
\includegraphics[
trim=-0.137272in 0.000000in -0.137272in 0.000000in,
height=6.2171in,
width=6.2163in
]%
{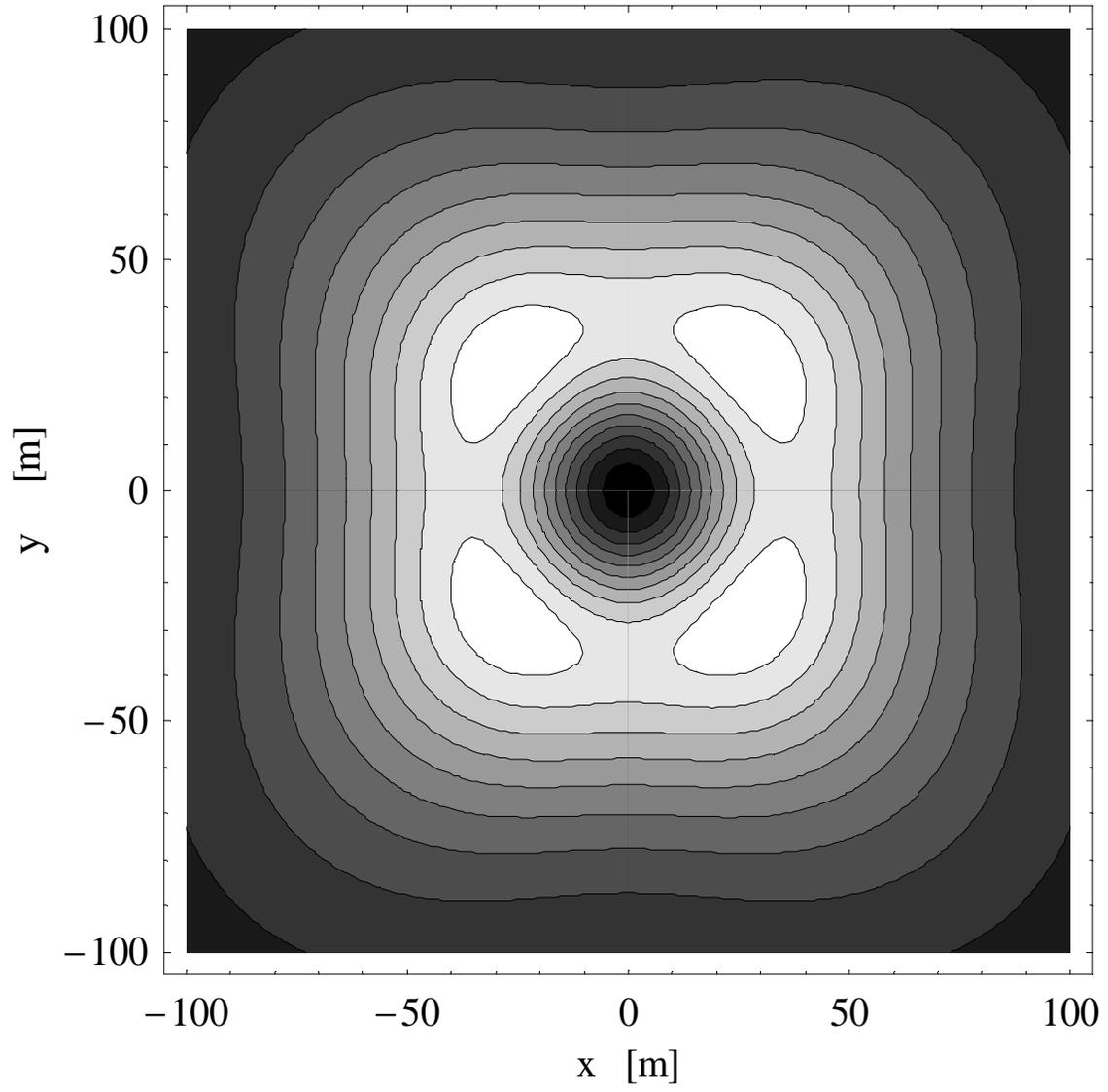}%
\caption{A plot of the contours of constant $1/R_{xyz}$ is shown in the
$x_{o}-y_{o}$ plane at $z_{o}=100/\sqrt{3}\operatorname{m}$. Light-shaded
areas are small values of $R_{xyz}$. \ Units on both axes are meters.}%
\label{CaseAContourPlot}%
\end{center}
\end{figure}

\begin{figure}
[ptbptbptb]
\begin{center}
\includegraphics[
trim=-0.115337in 0.029169in -0.285484in -0.029169in,
height=3.8579in,
width=6.4454in
]%
{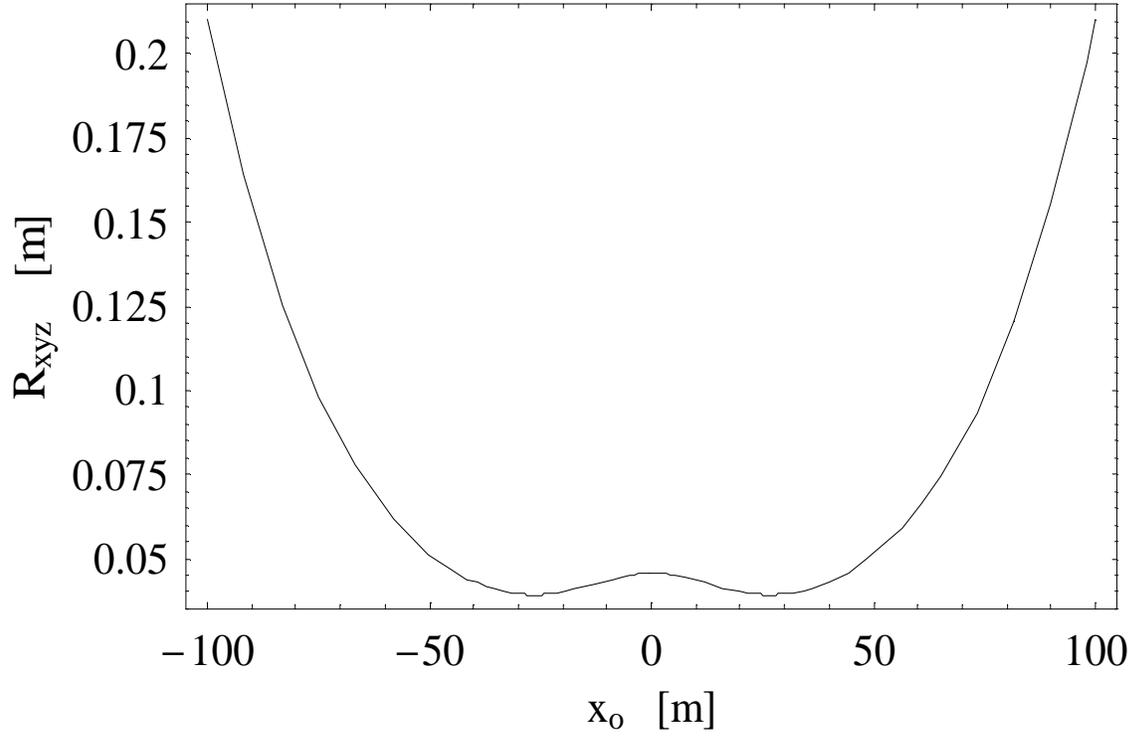}%
\caption{The error $R_{xyz}$ vs. $x_{o} $ is plotted for $y_{o}%
=30\operatorname{m}$ and $z_{o}=100\operatorname{m}$/$\sqrt{3} $, which
corresponds to a line in Figure 4 with relativily small error $R_{xyz}$. The
same parameters are used in this plot as in Figure 4. \ Units on both axes are
meters.}%
\label{CaseAPlot1}%
\end{center}
\end{figure}

\begin{figure}
[ptbptbptbptb]
\begin{center}
\includegraphics[
trim=-0.496684in 0.000000in -0.496684in 0.000000in,
height=4.2263in,
width=6.8208in
]%
{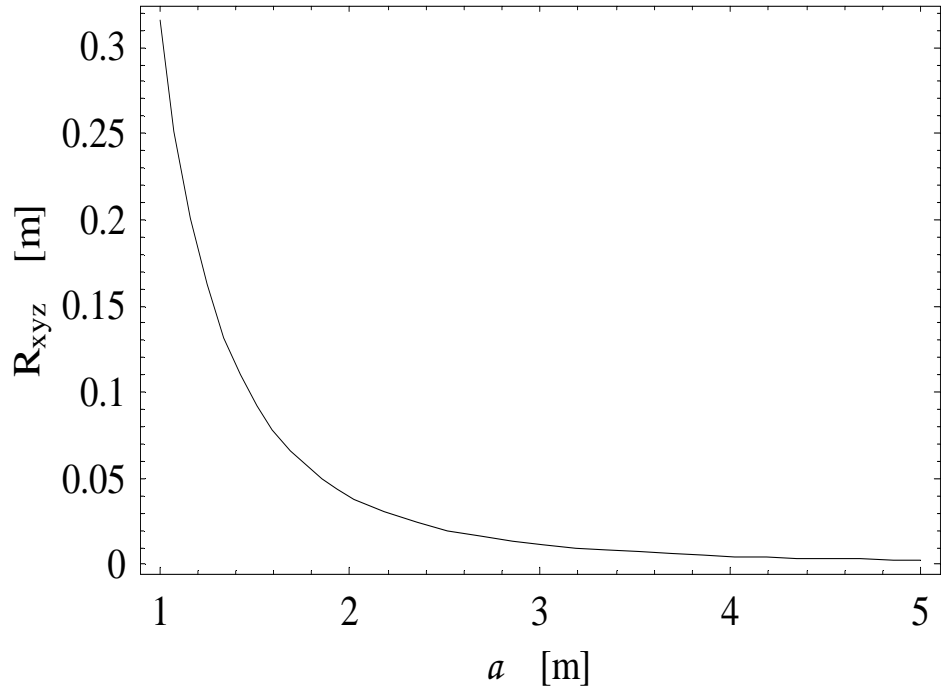}%
\caption{The error $R_{xyz}$ vs. $a$ (half the interferometer baseline length)
is plotted for $x_{o}=y_{o}=30\operatorname{m}$, and $z_{o}%
=100\operatorname{m}$/$\sqrt{3}$, which corresponds to the high-accuracy
light-shaded region in upper right of Figure 4. \ Units on both axes are
meters.}%
\label{CaseAPlot2}%
\end{center}
\end{figure}

\begin{figure}
[ptbptbptbptbptb]
\begin{center}
\includegraphics[
height=5.8237in,
width=5.2338in
]%
{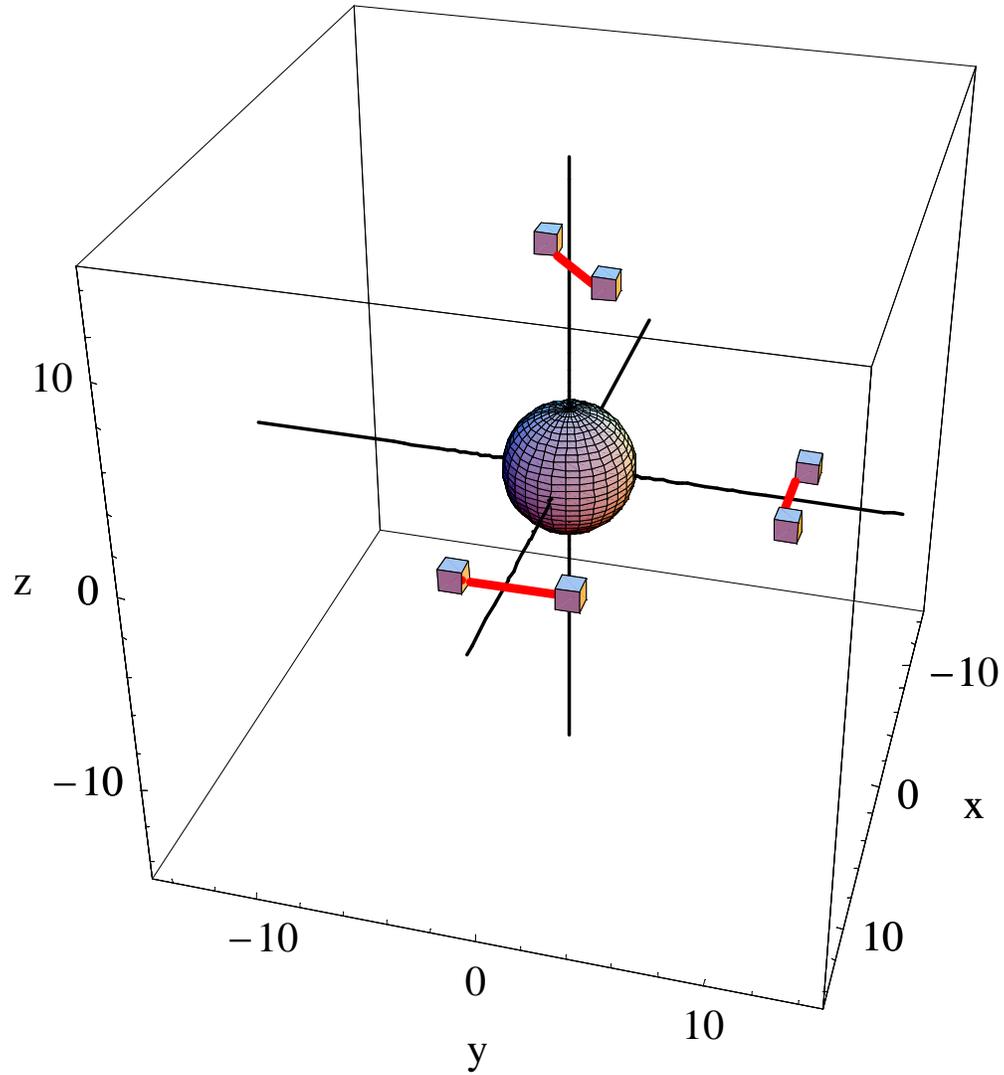}%
\caption{A schematic of the LEO satellite QPS is shown. \ Pairs of satellites
orbiting Earth, shown by connecting lines, form the interferometer baselines
of length $b$. \ Example numbers used in this calculation have baseline
$b=20\operatorname{km}$ and LEO satellite semi-major axis
$a=7360\operatorname{km}.$}%
\label{QPS plot}%
\end{center}
\end{figure}

\begin{figure}
[ptbptbptbptbptbptb]
\begin{center}
\includegraphics[
height=5.3074in,
width=5.3074in
]%
{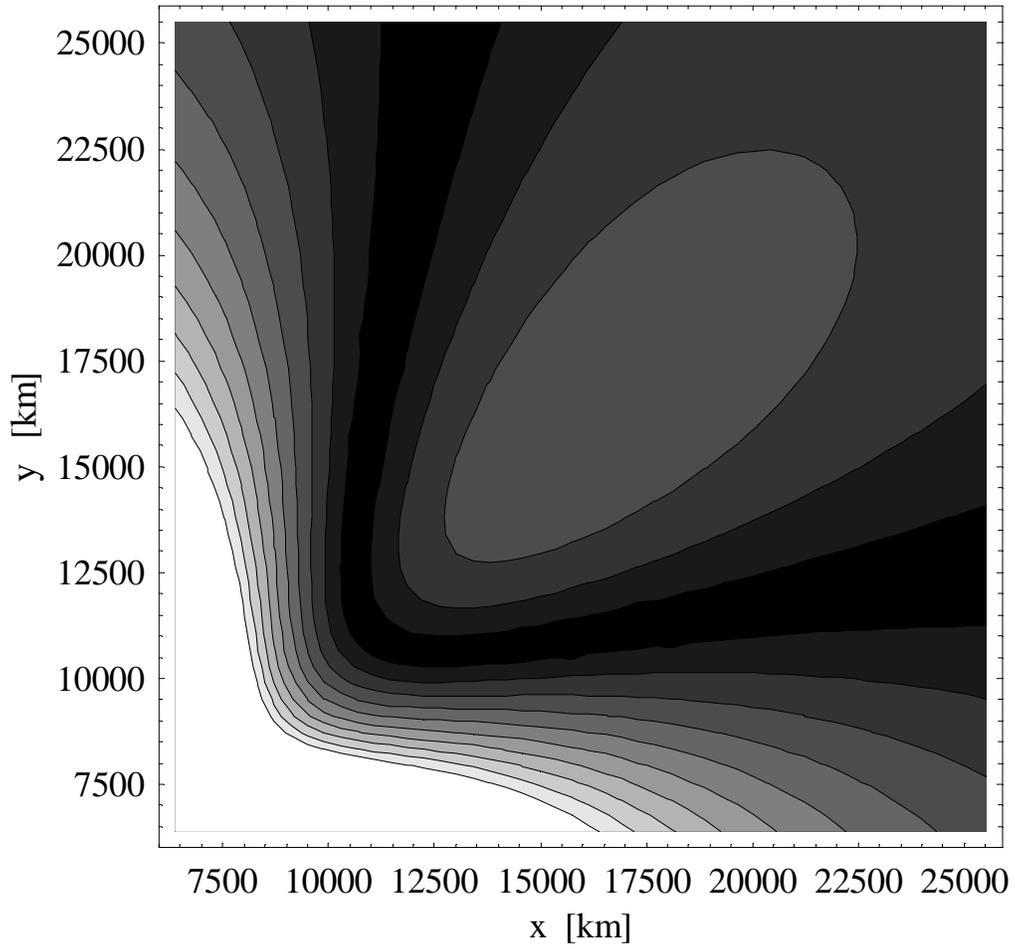}%
\caption{Contours of constant reciprocal position error, $1/R_{xyz}$, are
shown in the $x_{o}-y_{o}$ plane for $z_{o}=R_{e}/\sqrt{3}$\ and $\sigma
_{s}=1.0\operatorname{\mu m}$. Lighter-shaded areas are smaller values of
error $R_{xyz}$. \ The semi-major axis of the LEO satellites is taken to be
$a=7360\operatorname{km}$ and the baselines (satellite pair separation)
$b=20\operatorname{km}$. }%
\label{CaseCContourPlot}%
\end{center}
\end{figure}

\begin{figure}
[ptbptbptbptbptbptbptb]
\begin{center}
\includegraphics[
trim=-1.208424in 0.000000in -1.208424in 0.000000in,
height=4.3171in,
width=6.986in
]%
{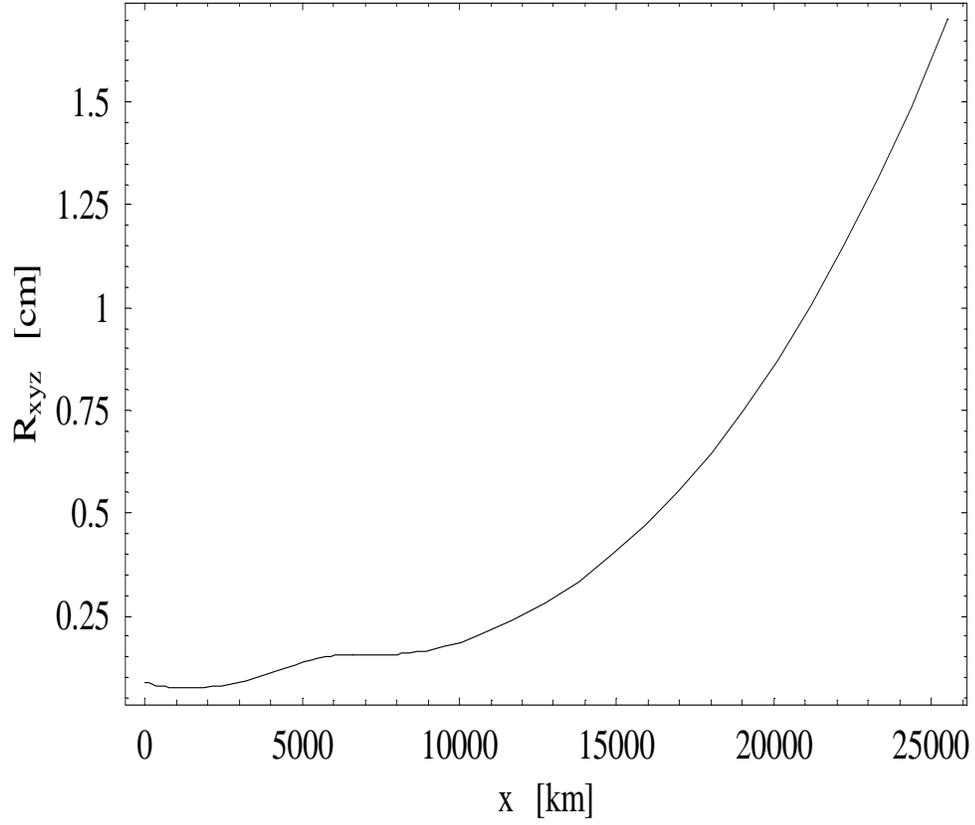}%
\caption{Plot of position error, $R_{xyz} $ vs. $x_{o}$, shown for
$y_{o}=z_{o}=R_{e}/\sqrt{3}$ with $\sigma_{s}=1.0\operatorname{\mu m}$. The
semi-major axis of the LEO satellites is taken to be $a=7360\operatorname{km}$
and the baselines (satellite pair separation) $b=20\operatorname{km}$. }%
\label{CaseCPlot1}%
\end{center}
\end{figure}

\begin{figure}
[ptbptbptbptbptbptbptbptb]
\begin{center}
\includegraphics[
trim=-0.710769in 0.000000in -0.710769in 0.000000in,
height=4.0819in,
width=6.5951in
]%
{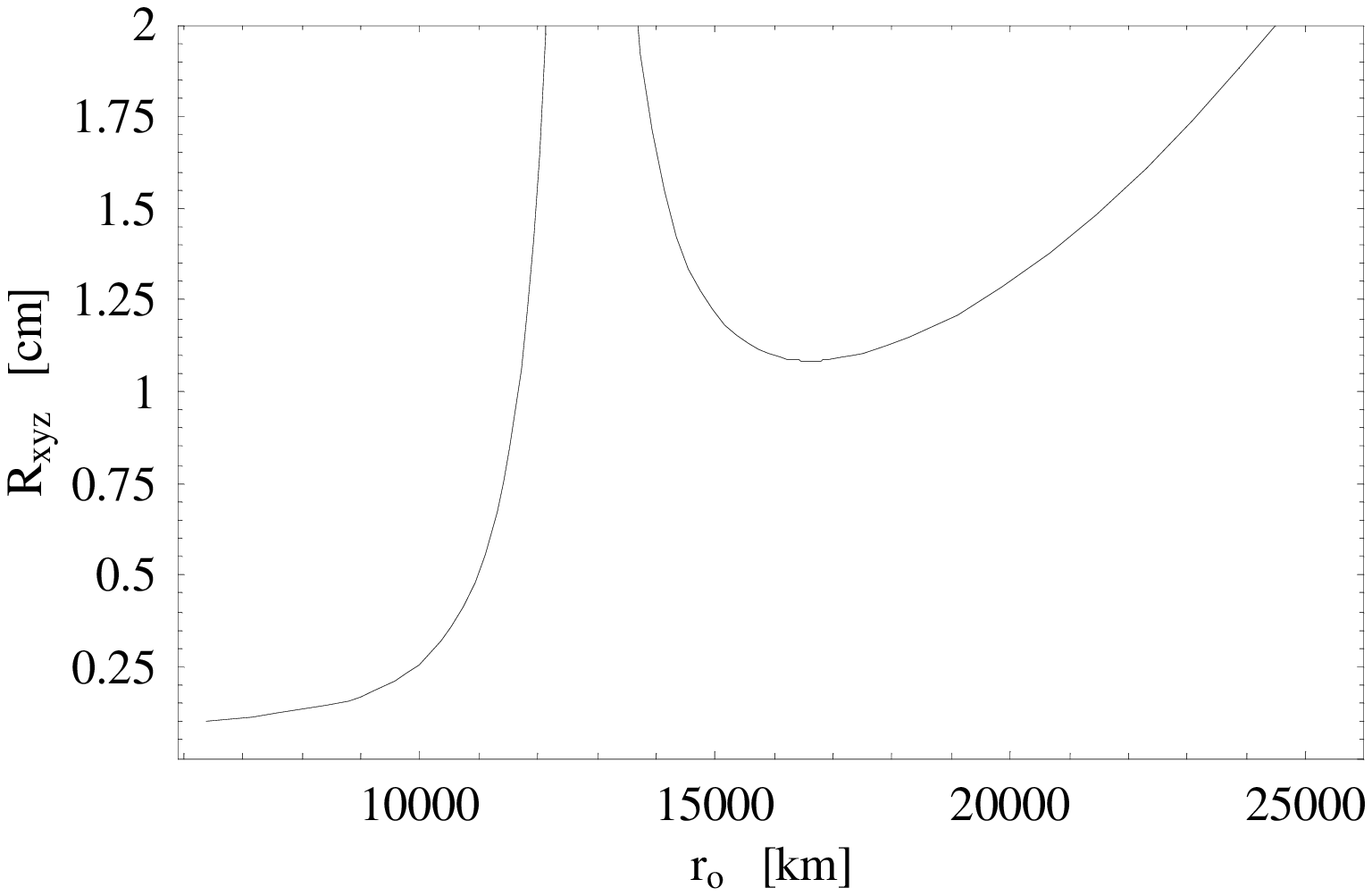}%
\caption{Plot of position error in the radial direction, $R_{xyz}$ vs. $r_{o}%
$, \ where $r_{o}=\sqrt{x_{o}^{2}+y_{o}^{2}+z_{o}^{2}}$ \ and $\sigma
_{s}=1.0\operatorname{\mu m}$. The semi-major axis of the LEO satellites is
taken to be $a=7360\operatorname{km}$ and the baselines (satellite pair
separation) $b=20\operatorname{km}$. }%
\label{CaseCPlot2}%
\end{center}
\end{figure}

The scenarios that we have described above are ones where the measurements
(adjusting the optical delays) and the calculations (to compute $\mathbf{r}%
_{o}$) are done near the origin of the reference frame. \ In a classical GPS
receiver, the computations (correlations of PRN\ codes to at least four
satellites) are done locally in the user's GPS receiver that the user carries
with him. The QPS analog of this classical GPS scenario is a setup where the
biphotons are generated at points $E_{1\text{,}}$ $E_{2}$, and $E_{3}$, but
the user carries with him the 50:50 beam splitters and photon detectors. In
this scenario, the user controls (and carries with him) the optical delays,
see Figure \ref{baselineFigure}, and he locally measures the optical delays
$s_{1\text{, }}s_{2}$, and $s_{3}$. The user must receive a classical message
consisting of the coordinates of baseline endpoints, $\mathbf{R}_{i}$,
$i=1,...,6$, and then he must solve the Eq.(\ref{baseline 1}%
)-(\ref{baseline 3}) for his position $\mathbf{r}_{o}=(x_{o},y_{o},z_{o})$. In
such a case, there are no clocks on-board the broadcasting satellites (located
at positions $\mathbf{R}_{i}$), however, the user must carry a clock with
short term stability to determine rate of photon coincidence counts from each
of the three baselines (associated with spatial positioning) and also he must
do coincidence counting for time determination (if time is needed). \ For the
three spatial baselines, optical propagation is then one-way (using the
satellite positions as a primary reference system, see below) from satellites
to QPS\ user receiver. \ For time synchronization, however, as mentioned
previously, the optical propagation must be two-way (when using the method of
Bahder and Golding). In essence, for each of the four channels, the QPS
receiver consists of a beam splitter, two single-photon detectors, and a
controllable optical delay. All four space-time coordinates can be obtained by
a user in this way. One clock in the reference frame must have long-term
stability to define coordinate time, and another clock in the QPS user
receiver can have short-term stability. \ \textit{Note that the satellites do
not need to carry clocks, because their positions can be used to define the
primary system of coordinates. }\ This type of QPS is a close analog of the
classical GPS.

\section{QPS\ Space-time Coordinates}

The satellites at baseline points $\mathbf{R}_{i}$\ can be taken to define the
\textit{primary} system of reference, even though the points $\mathbf{R}_{i}$
change with time. The quantities measured by a user of such a QPS are then
$(s_{0},s_{1},s_{2},s_{3})$, where $s_{0}$ is the optical time delay (in the
HOM\ interferometer) that will provide the user with coordinate time in this
coordinate system (using the Bahder and Golding method), and $(s_{1}%
,s_{2},s_{3})$ are the three optical delays in the three interferometers for
position determination. \ The quantities $(s_{0},s_{1},s_{2},s_{3})$ are then
to be regarded as generalized 4-dimensional space-time
coordinates\cite{Synge1960}, $s_{0}$\ is a time-like coordinate and
$(s_{1},s_{2},s_{3})$\ \ are space-like coordinates. \ Within the context of
general relativity, such coordinates are as good as any other coordinates, and
they enter into the metric $c^{2}d\tau^{2}=g_{ij}ds_{i}ds_{j}$\ of the flat
space-time assumed in this work. \ Of course, a transformation from the QPS
space-time coordinates, $(s_{0},s_{1},s_{2},s_{3})$, to an Earth-centered
inertial (ECI)\ system of coordinates, say $(t,x,y,z)$, is of interest for
astrodynamic applications. \ \ Such a transformation can be done approximately
by conventional means of tracking the satellites (at baseline points
$\mathbf{R}_{i}$).

It is interesting to remark that the QPS allows the direct measurement of
4-dimensional space-time coordinates. \ Previously, it was believed that
space-time coordinates were not measurable
quantities\cite{Synge1960,Pirani1957,Soffel1989,Brumberg1991}. \ Of course,
the QPS\ coordinates $(s_{0},s_{1},s_{2},s_{3})$ are real physical
measurements, and it is well-known that real measurements are space-time
invariants under generalized coordinate transformations\cite{Synge1960}.

\section{Summary}

I have presented a conceptual scheme for an interferometric quantum
positioning system (QPS)\ based on second order quantum coherence of entangled
photon pairs (biphotons). A user's spatial coordinates are determined by
locating three unique minima in three different two-photon counting rates,
associated with three HOM\ interferometers built on independent baselines.
\ The spatial portion of the QPS\ is similar to a classical TOA system,
however, a classical TOA system requires synchronized clocks that keep
coordinate time, which is often a difficult requirement to meet. \ In
contrast, the QPS only requires a clock having a short-term stability to
measure two-photon coincidence counting rates while the optical time delay is
adjusted (to locate the minima in the two-photon coincidence counting rate
$R_{c}$). Bright sources of entangled photons (biphotons) are needed.

Several different scenarios were considered for a QPS: one is a terrestrial
system\ and the another is space-based. \ In both cases, I computed the
accuracy of a user's position as a function user position. The function that
describes the errors in position has a complex spatial dependence. \ In the
case of the terrestrial QPS, the position accuracy was relatively poor because
the baselines were located near each other. This could be dramatically
improved by moving apart the baselines.

As an example of a satellite-based QPS, I have proposed a LEO-satellite QPS.
\ Neglecting atmospheric effects, calculations suggest that the position
accuracy $R_{xyz}$ of such a QPS\ can be below the 1 cm-level for an error
(standard deviation) in the optical delays $\sigma_{s}=1.0%
\operatorname{\mu m}%
$ associated with the minima in two-photon counting rates $R_{c}$. \ The
complex dependence of $R_{xyz}$\ on user position suggests that significant
engineering must be done to design a realistic QPS.

\begin{acknowledgments}
This work was sponsored by the Advanced Research Development Activity (ARDA).
\ The author is grateful to Yanhua Shih for helpful communications.
\end{acknowledgments}

\newpage%

\newpage%

\begin{thebibliography}{99}                                                                                               %
\bibitem {ParkinsonSpilker1996}\textit{Global Positioning System: Theory and
Applications}, Vols. I and II, edited by B. W. Parkinson and J. J. Spilker,
Progress in Astronautics and Aeronautics, Vols. 163 and 164 Amer. Inst. Aero.
Astro., Washington, D.C., 1996.

\bibitem {Kaplan96}E. D. Kaplan, Understanding GPS: Principles and
Applications, Mobile Communications Series (Artech House, Boston, 1996).

\bibitem {Hofmann-Wellenhof93}B. Hofmann-Wellenhof, H. Lichtenegger, and J.
Collins, Global Positioning System Theory and Practice (Springer-Verlag, New
York, 1993).

\bibitem {Bahder2001}T. B. Bahder, Am. J. Phys. \textbf{69}, 315, 2001,
"Navigation in curved space-time".

\bibitem {Bahder2003}T. B. Bahder, \ Phys. Rev. D \textbf{68}, 063005 (2003).

\bibitem {Chuang2000}I. L. Chuang, Phys. Rev. Lett. \textbf{85}, 2006 (2000),
and also in quant-ph/0004105.

\bibitem {Jozsa2000}R. Jozsa, D. S. Abrams, J. P. Dowling, and C. P. Williams,
Phys. Rev. Lett. \textbf{85}, 2006 (2000).

\bibitem {Giovannetti2001}V. Giovannetti, S. Lloyd, and L. Maccone, Nature
\textbf{412}, 417 (2001).

\bibitem {Giovannetti2001b}V. Giovannetti, S. Lloyd, and L. Maccone, F. N. C.
Wong, Phys. Rev. Lett. 87, 117902 (2001).

\bibitem {Giovannetti2002}V. Giovannetti, S. Lloyd, and L. Maccone, Phys. Rev.
A, \textbf{65}, 022309 (2002).

\bibitem {Shih2003}Y. Shih, unpublished.

\bibitem {Bahder2004}T. B. Bahder and W. M. Golding, \textquotedblleft Clock
synchronization based on second-order coherence of entangled photons",
submitted for publication in Phys. Rev. A , April 2004.

\bibitem {Bahder2004b}T. B. Bahder, "Clock Synchronization and Navigation in
the Vicinity of the Earth", gr-qc/0405001, to be published in "Progress in
General Relativity and Quantum Cosmology Research", Nova Science Publishers,
Inc., Hauppauge, New York, December 2004.

\bibitem {Burnham1970}D. C. Burnham and D. L. Weinberg, Phys. Rev. Lett.
\textbf{25}, 84 (1970).

\bibitem {HOM1987}C. K. Hong, Z. Y. Ou, and L. Mandel, Phys. Rev. Lett.
\textbf{59}, 2044 (1987).

\bibitem {Ghosh1986}R. Ghosh, C. K. Hong, Z. Y. Ou, and L. Mandel, Phys. Rev.
A \textbf{34}, 3962 (1986).

\bibitem {KlyshkoBook1988}D. N. Klyshko, \textit{Photons and Nonlinear Optics}
(Gordon \& Breach, New York, 1988).

\bibitem {Rubin1994}M. H. Rubin, D. N. Klyshko, Y. H. Shih, and A. V.
Sergienko, Phys. Rev. A \textbf{50}, 5122 (1994).

\bibitem {delaytime}This optical time delay can be a free-space delay, or it
may be a high-index optical material medium, such as cold atoms, see for
example: C. Liu, Z. Dutton, C. H. Behroozi, and L. V. Hau, "Observation of
coherent optical information storage in an atomic medium using halted light
pulses", Nature \textbf{409}, 490 (2001).

\bibitem {Glauber1965}R. J. Glauber, \textit{Optical Coherence and Photon
Statistics}, Lectures delivered at Les Houches during 1964 Summer School of
Theoretical Physics at University of Grenoble, Gordon Breach Science
Publishers, New York (1965).

\bibitem {Glauber1963}R. J. Glauber, Phys. Rev. \textbf{130}, 2529 (1963);
Phys. Rev. \textbf{131}, 2766 (1963)

\bibitem {Mandel1995}L. Mandel and E. Wolf, "\textit{Optical Coherence and
Quantum Optics}", Cambridge Iniversity Press, New York (1995).

\bibitem {clock stability}For example, the GPS satellite clocks (essentially)
keep coordinate time and must have a good long-term stability, see
Ref.\cite{Bahder2003}.

\bibitem {Bevington2003}P. R. Bevington and D. K. Robinson, "\textit{Data
Rediction for the Physical Sciences}" McGraw-Hill, New York, (2003).

\bibitem {Bahder1995Rep}T. B. Bahder, "GPS for Land Combat Applications",
Summary Report of the Army Workshop held at University of North Carolina, 2
and 3 August 1995, \ Army Research Laboratory Technical Report ARL-SR-40,
December 1995.

\bibitem {Grice1997}W. P. Grice and I. A. Walmsley, Phys. Rev. A, \textbf{56},
1627 (1997).

\bibitem {Hughes2002}R. J. Hughes, J. E. Nordholt, D. Derkacs and C. G.
Peterson, New Journal of Physics \textbf{4, }43.1 (2002).

\bibitem {Rarity2002}J. G. Rarity, P. R. Tapster, P. M. Gorman and P. Knight,
New Journal of Physics \textbf{4, }82.1 (2002).

\bibitem {Steinberg1992}A. M. Steinberg, P. Kwiat, and R. Y. Chiao, Phys. Rev.
A \textbf{45}, 6659 (1992).

\bibitem {Steinberg1992a}A. M. Steinberg, P. Kwiat, and R. Y. Chiao, Phys.
Rev. Lett. \textbf{68}, 2421 (1992).

\bibitem {Perina1999}J, Perina, A. Sergenko, B. M. Jost, B. E. A. Saleh, and
M. Teich, Phys. Rev. A \textbf{59}, 2359 (1999).

\bibitem {Valencia2002}A. Valencia, M. Chekhova, A. Trifonov, and Y. Shih,
Phys. Rev. Lett. \textbf{88}, 183601 (2002).

\bibitem {Synge1960}J. L. Synge, \textit{Relativity:\ The General Theory},
North-Holland, New York (1960).

\bibitem {Pirani1957}A. E. Pirani, Bulletin De L'Academie Polonaise Des
Sciences Cl. III, 1957, Vol. V, No. 2, p. 143-146 (1957).

\bibitem {Soffel1989}M. H. Soffel, \textit{Relativity in Astrometry, Celestial
Mechanics and Geodesy}, Ch. 3, Springer-Verlag, New York (1989).

\bibitem {Brumberg1991}V. A. Brumberg, In section 2.3 of \textit{Essential
Relativistic Celestial Mechanics}, published under the Adam Hilger imprint by
IOP Publishing Ltd, Techno House, Bristol, England (1991).

\bibitem {Guinot1997}B. Guinot, Application of general relativity to
metrology", Metrologia \textbf{34}, \ 261 (1997).\newpage
\end{thebibliography}
\end{document}